\documentclass[superscriptaddress,showpacs,showkeys,twocolumn,prl]{revtex4}
\usepackage{graphicx}
\usepackage{amsmath}

\begin{document}

\title{Nanoscale periodicity in stripe-forming systems at high temperature: 
Au/W(110)}
\author{J. de la Figuera}
\affiliation{Insituto de Qu\'{\i}mica-F\'{\i}sica ``Rocasolano'', CSIC, Madrid 28006 Spain}
\affiliation{Universidad Aut\'{o}noma de Madrid, Madrid 28049, Spain}
\author{F. L\'{e}onard}
\affiliation{Sandia National Laboratories, Livermore, California 94550, USA}
\author{N. C. Bartelt}
\affiliation{Sandia National Laboratories, Livermore, California 94550, USA}
\author{R. Stumpf}
\affiliation{Sandia National Laboratories, Livermore, California 94550, USA}
\date{\today}
\author{K. F. McCarty}
\affiliation{Sandia National Laboratories, Livermore, California 94550, USA}
\date{\today}

\begin{abstract}
We observe using low-energy electron microscopy the
self-assembly of monolayer-thick stripes of Au on W(110) near
the transition temperature between stripes and the non-patterned (homogeneous) phase. We
demonstrate that the amplitude of this Au stripe phase decreases with 
increasing temperature and vanishes at the order-disorder
transition (ODT). The wavelength varies much more slowly with temperature and coverage
than theories of stress-domain patterns with sharp phase boundaries would predict, and
maintains a finite value of about 100 nm at the ODT.  We argue that such nanometer-scale stripes should often appear near the ODT. \end{abstract}


\maketitle
The theory that ordered patterns can occur on solid surfaces to relax stress was proposed many years ago\cite{Marchenko,Alerhand88}.  These patterns in principle offer a way of controlling the structure and hence functionality of surfaces.
 ``Stress-domain'' patterns arise from the competition between the short-range attractive
interaction between atoms, leading to a phase-boundary energy, and a
long-range repulsive interaction between  boundaries, due to the difference
in surface stress between the two phases. This repulsion is
mediated by elastic deformations of the substrate.    So far, such stress-domain
patterns have been observed and
quantified in the low-temperature, sharp-interface regime, where the
interfaces between the two separated phases are abrupt. (See for example \cite{CuO,Plass0,hannon}.) However, as the temperature is increased,  the amplitude of
the modulated pattern should decrease.  At sufficiently high temperature, a
transition to a homogeneous phase occurs. We call this temperature the order-disorder
transition (ODT). As the ODT is approached, the interface width is expected to
increase, eventually becoming of the order of the stripe periodicity,  making
the sharp-boundary theory\cite{Alerhand88,Marchenko} inappropriate.

Our present experimental study of Au on W(110) explicitly shows this breakdown near an ODT. We also show that the breakdown of the sharp-boundary limit has a large consequence on the pattern periodicity. The Au-stripe periodicity varies much more slowly with temperature and coverage
than theories of stress-domain patterns with sharp phase boundaries would predict, and
maintains a finite value of about 100 nm at the ODT.  In contrast, we find that the mean-field description of patterns with diffuse interfaces well-describes the measured temperature dependence. Based on this agreement, we suggest that nanometer-scale periodicities should be much more
common in pattern-forming systems at their high-temperature limit than one would expect from the low-T, sharp-interface theory.

\begin{figure}[th]
\caption{a) LEEM image of Au stripes, which appear dark at 10 eV electron energy, on W(110) at 619$%
{{}^\circ}%
$C. The field of view is 7~$\protect\mu $m. b) Measured phase diagram of Au on
W(110). The solid line is a guide to the eye to show when the Au adatom gas and the Au islands
are in equilibrium. Filled circles are data from reflectivity
measurements\cite{Ag_SS2006}. Open circles are determined from loss of contrast in LEEM
images. Stripes were observed at temperatures near but below the open
circles.}
\label{phase}\centering \includegraphics[width=0.5\textwidth]{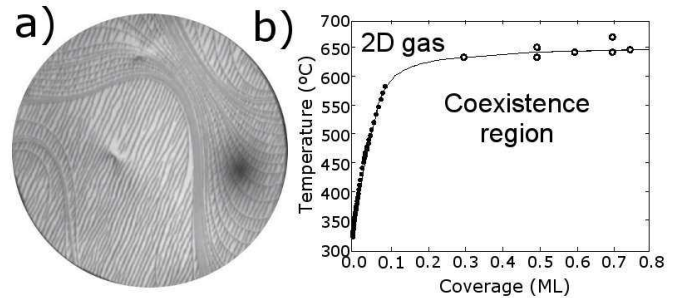}
\end{figure}

Experimentally, quantitative observations near the ODT are difficult because thermal
fluctuations of boundaries typically destroy the pattern's long-range order. Here we study stripe formation with long-range order in the system of Au on
W(110). As first observed by Duden and Bauer\cite{DudenTh,bookBauer},
submonolayers of Au on W(110) self-assemble into stripe patterns,
which consist of monolayer-thick stripes of condensed-phase Au in
coexistence with stripes of a Au adatom gas (see Fig.~\ref{phase}a). Because
of strong surface anisotropy, the stripes form along a
particular crystallographic direction, [1$\overline{1}$0], and we are able to use low-energy
electron microscopy (LEEM) to measure the pattern's amplitude (related
to the Au density\cite{Ag_SS2006}) and wavelength approaching the ODT.
The amplitude decreases steadily with increasing
temperature and vanishes at the ODT. The pattern's wavelength also
decreases with temperature but has a finite value of 100 nm at the ODT.

The W(110) single-crystal substrate cleaning and noble-metal growth are
described in detail elsewhere\cite{Ag_SS2006}. Our estimated error in the
absolute temperature is 10 K with a precision of
1 K. The patterns in the Au films were imaged by LEEM both during Au deposition and
as a function of temperature. The total Au coverages, given as the fraction of a complete condensed Au layer in monolayers (ML), were determined by the elapsed deposition time from a calibrated Au flux.

We first establish the location of the
pattern formation in the Au on W(110) phase diagram. Small amounts of Au form a two-dimensional adatom gas. At higher densities, the Au adatoms
condense into monolayer-thick Au islands. Because Au and W do not
alloy\cite{alloybook},
the system has long been used to evaluate phase diagrams of condensates and
adatom gases\cite{KOLACZKIEWICZ1985,KOLACZKIEWICZ1985a}. Fig.~\ref{phase}b shows
the
phase diagram measured using LEEM. At low temperatures, we determine the Au
adatom concentration in equilibrium with condensed Au using the linear decrease in electron reflectivity with coverage \cite{Ag_SS2006}. As seen in
Fig.~\ref{phase}b (filled circles), 
the adatom density in equilibrium with condensed Au increases with temperature. 
At higher temperatures and coverages, we find
that the system self-assembles into a lamellar phase consisting of stripes
of condensed Au separated by stripes of the Au adatom gas (Fig.~\ref{phase}a).

\begin{figure}[th]
\centering \includegraphics[width=0.5\textwidth]{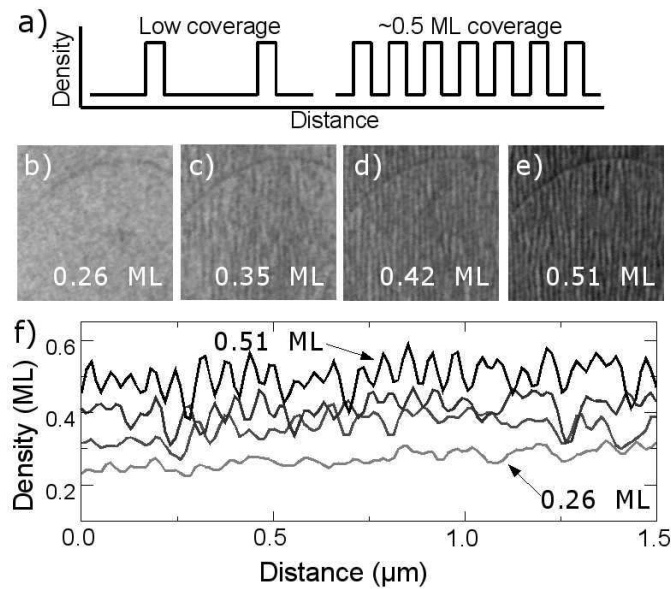}
\caption{a) Schematic of the expected changes in the stripe pattern with coverage in the sharp-interface limit. In contrast, b)-e) show LEEM images at constant temperature (620$^\circ$C) while continuously depositing Au on the W(110) surface. The images are
 4~$\protect\mu $m wide. All frames use the same intensity scale.
 f) Representative intensity profiles (averaging 10 lines of pixels for 1 second) at each coverage.}
\label{coverage}
\end{figure}

In the sharp-interface picture of stripe formation, the stripes can be considered as being composed of two coexisting phases.   In this limit, the density of
the Au-adatom-gas region of the stripes as well as the density of the condensed Au islands 
at any temperature should be that given
by the phase diagram partially shown in Fig.~\ref{phase}b. The relative areas of the two phases should be determined
 by the lever rule (see schematic of Fig.~\ref{coverage}a). In particular, at constant temperature the densities of both phases should be independent of the total Au coverage.

As shown in Fig.~\ref{coverage}, our observations are inconsistent with the sharp-interface limit.  Fig.~\ref{coverage}(b)-(e) shows images when the total
Au coverage is varied from 0.2 ML to 0.5 ML. The uniform
decrease in image intensity with increasing coverage indicates\cite{Ag_SS2006} that the atomic
density of both the dilute and dense regions of the stripes is changing with
Au coverage.  The changing density of the stripes is  shown
more directly in Fig.~\ref{coverage}(f), which gives profiles across these
images where we have converted image intensity to Au density, following \cite{Ag_SS2006}. Rather than the relative area of the two phases changing with
overall coverage, the relative area is constant and  the density of both phases
changes. For example, the density of the gas-phase Au regions changes from about 0.25 ML to 0.4 ML when  the total
Au coverage is varied from 0.2 ML to 0.5 ML.

\begin{figure}[th]
\centering \includegraphics[width=0.5\textwidth]{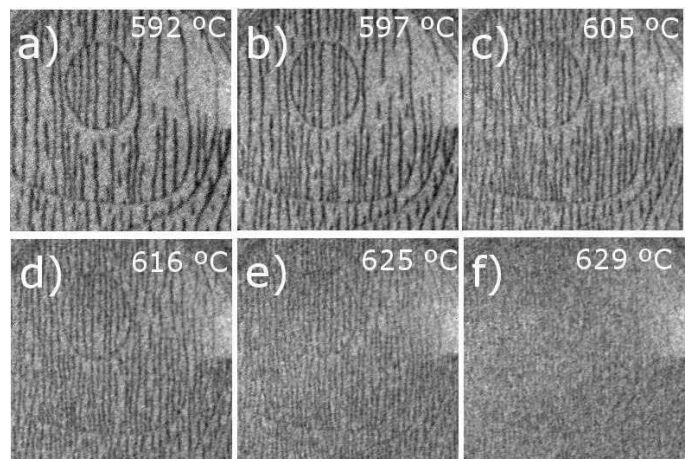}
\caption{a)-f) Sequence of LEEM images taken while changing the temperature
at a constant Au coverage. The images are
4~$\protect\mu $m wide.}
\label{temperature}
\end{figure}

\begin{figure}[th]
\centering \includegraphics[width=0.5\textwidth]{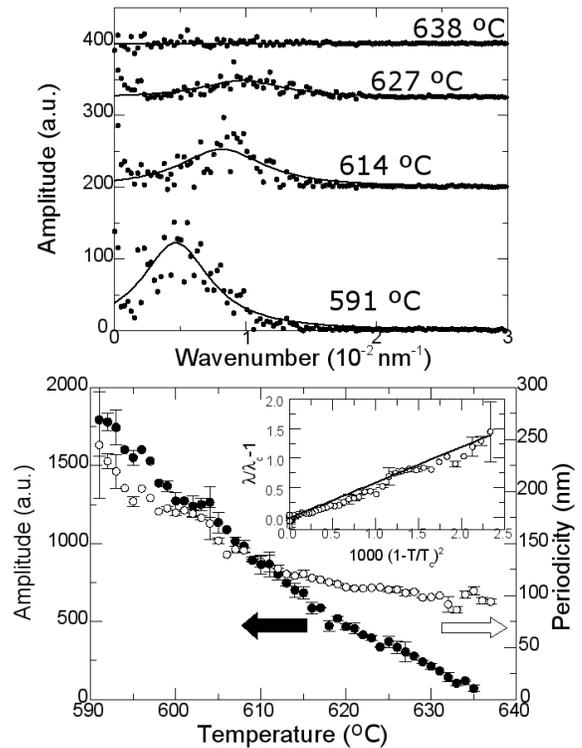}
\caption{a) Plot of the Fourier spectra, offset for clarity, along the x-direction at different temperatures. b) Temperature dependence of the stripe amplitude
and periodicity. Inset: Stripe periodicity
showing the scaling as described in Eq. (4).}
\label{dependence}
\end{figure}

We next show that the pattern's wavelength and amplitude as temperature approaches the ODT are markedly different from the behavior expected from the sharp-interface limit.
As seen in Fig.~\ref{temperature}, with increasing temperature the stripe density 
increases and the pattern contrast decreases. 
To quantify the amplitude and wavelength of the stripe phase, we used a Fourier-transform approach\cite{FFT}.
Results of this procedure applied to
a sequence of images at different temperatures are shown in
Fig.~\ref{dependence}. As
temperature increases, the amplitude of the pattern decreases steadily
until it vanishes at the ODT. While the periodicity of the pattern also
initially decreases with increasing temperature, it does not approach zero
at the ODT. Instead, the wavelength reaches a constant value of about 100
nm.

This striking coverage and temperature dependence and the fact that the stripes form near temperatures where the gas-phase and condensed Au become indistinguishable suggest that the diffuse-interface limit of stress-domain patterns is appropriate. To explain the observed periodicity in terms of atomic parameters, we adopt a continuum diffuse-interface model with a
spatially varying order parameter $\phi =\left( 2\rho -\rho _{0}\right)
/\rho _{0}$, where $\rho $  and $\rho _{0}$ are the densities, respectively, of the local Au and the condensed Au islands at low temperature. The free energy of the system is a sum
of short-range and long-range contributions, $F=F_{sr}+F_{lr}$. The
short-range energy per unit stripe length is %
\begin{equation}
F_{sr}=\int dx\left[ -\frac{r}{2}\phi ^{2}+\frac{u}{4}\phi ^{4}+\frac{c}{2}%
\left( \frac{d\phi }{dx}\right) ^{2}\right] .
\end{equation}%
Here $x$ is the coordinate perpendicular to the stripes, $%
r=r_{0}(T_{c}^{0}-T)$ with $T_{c}^{0}$ the bare transition temperature
(i.e., before renormalization by the long-range interaction), $r_{0}$ and 
$u$ are phenomenological parameters and $c$ determines the boundary energy.
This form for the short-range free energy is the standard Landau
expression representing a phase diagram with a low-temperature miscibility
gap and a critical point at $T_{c}^{0}$. 

The long-range interaction comes from the elastic fields in the substrate
due to the presence of surface atomic stress dipoles. We assume that the
local surface stress $\tau$ is proportional to the adatom concentration: $\tau =\phi
\Delta \tau /2$, with $\Delta \tau =\rho _{0}\left( \partial \tau /\partial
\rho \right)$.
Then the total long-range energy per unit stripe length is%
\begin{equation}
F_{lr}=2g\int dxdx^{\prime }\frac{\phi (x)\phi (x^{\prime })}{\left(
x-x^{\prime }\right) ^{2}+a^{2}}.
\end{equation}%
Here $g=\left( \Delta \tau \right) ^{2}M/2\pi $, where $M$ is a combination
of elastic constants\cite{Leonard2005}, and $a$ is on the order of an atomic
lattice constant. Far from the ODT, $\phi =\pm 1$, with sharp interfaces
separating the stripes. In this regime, the equilibrium wavelength $\lambda $
is obtained from the above free energy by standard methods\cite%
{Alerhand88,Marchenko}: $\lambda =2\pi ae^{\left( 1+\frac{\beta }{g}\right) }
$, where $\beta $ is the boundary energy in the sharp-boundary limit. Near
the ODT, however, the amplitude of the modulation is small, and a
single-mode analysis is appropriate. Assuming a profile for the order
parameter $\phi (x)=A\cos (2\pi x/\lambda )$, and minimizing the free energy
as a function of $\lambda $, we find that at the renormalized ODT temperature, $T_{c}=T_{c}^{0}-g/\left( r_{0}a\right)$ \cite{McConnel991,Gao2003}:
\begin{equation}
\lambda _{c}=\frac{2c}{g}.  \label{lambdac}
\end{equation}%
This wavelength does not depend on coverage, exactly as we observe in Fig.~\ref{coverage}\cite{amplitude}.

In contrast to the sharp-boundary approach, with its exponential dependence on
material parameters, the wavelength at the ODT has
a much milder linear dependence on the ratio of boundary to elastic energies.  Although we have neglected fluctuations in the stripes in deriving this equation, we expect the proportionality to $c/g$ to be independent of this assumption:  On dimensional grounds $c/g$ defines a length scale that sets a lower limit for the periodicity of stress domains at high temperature.

\begin{figure}[th]
\centering \includegraphics[width=0.4\textwidth]{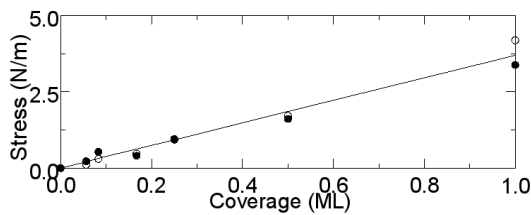}
\caption{Excess stress of Au adatoms on W(110) obtained from DFT
calculations. (Open circles, GGA; filled circles, LDA). }
\label{stress}
\end{figure}

To determine if Eq.~\ref{lambdac} quantitatively explains the observed periodicity, we must estimate the parameters $g$ and $c$.  
To calculate $g$, values for
the stress difference between the two phases and the constant $M$ are
needed. Fortunately, previous work on stripe phases at solid surfaces has
calculated the value of $M=4.5\times 10^{-12}$ m$^{2}$/N for stripes
oriented along the [1$\overline{1}$0] direction on the W(110) surface\cite{Leonard2005}.
To provide a value for the difference of surface stress between the
condensed-phase Au stripes and the Au adatom phase, we performed density
functional theory (DFT) calculations within the Local Density
Approximation (LDA) and the generalized gradient approximation (GGA)
\cite{PBE} of the excess stress of Au adatoms on W(110). The difference
between LDA and GGA values is an estimate of the unknown errors
introduced by the approximations to DFT. We use the VASP
code\cite{VASPAW} for our supercell calculations\cite{abinitio}. As
shown in Fig.~\ref{stress}, the calculated surface stress perpendicular to the stripe
direction depends linearly on the adatom coverage, providing the value $%
\Delta \tau =$ 3.7~N/m. The value for $g$ is thus $9.8\times 10^{-12}$ J/m.

Estimating $c$ is more difficult because it depends on the detailed nature of the short-ranged attractions between Au adatoms.   In general one expects $c$ to scale with $T_c$.  For example, in the simple nearest-neighbor Ising model on a square lattice $c = kT_c/2$.   To make a more refined estimate we consider the temperature dependence of the stripe
periodicity $\lambda $ close to the ODT. To do so, we extend the
single-mode analysis to two modes, and apply a perturbation analysis\cite{Leo97} to
calculate the amplitudes of the two modes at equilibrium.
Minimizing the free energy with respect to the wavelength using the
calculated amplitudes gives the stripe periodicity:
\begin{equation}
\lambda =\lambda _{c}\left[ 1+\frac{\lambda _{c}^{4}a^{4}r_{0}^{2}T_{c}^{2}}{%
384\pi ^{4}c^{2}}\left( 1-\frac{T}{T_{c}}\right) ^{2}\right] .
\label{lambda}
\end{equation}%
A plot of $\lambda /\lambda _{c}-1$ versus $\left( 1-T/T_{c}\right) ^{2}$
should give a straight line. Using the measured values $\lambda _{c}=100$ nm
and $T_{c}=908$ K, the inset in Fig.~\ref{dependence} shows that our experimental data follows the
prediction of Eq. (\ref{lambda}).  A numerical fit gives the ratio $a^{2}r_{0}T_{c}/c\approx 0.5$. From the regular solution
model\cite{Cahn}, we have $a^{2}r_{0}=4k_{B}$ giving $c\approx 8k_{B}T_{c}$. 
With this $c$ value and the $g$ given above, the stripe periodicity at the ODT predicted by Eq. $\left( \ref{lambdac}%
\right)\ $is then 20~nm.
Given our approximations, this number is in reasonable agreement with the
experimental result, $\lambda _{c}=$100~nm.

Although accurately calculating values of $c$ and $g$ in any particular system is difficult, there is nothing particularly special about Au/W(110).  $T_c$ is typical for metal monolayers, as is the calculated surface stress of the condensed overlayer. Thus we expect that the equilibrium structure of metal monolayers at high temperature will often have density modulations with a similar length scale. This is in striking contrast to the sharp-interface limit of stress domains, where the domain periodicity is extremely sensitive to interaction strengths.

In summary, we have presented experimental evidence that Au stripes observed on W(110) at high temperature are in the diffuse-interface limit of  surface-stress domains, with a temperature and coverage dependence qualitatively different from the oft-applied sharp-interface limit. By comparing our results with theoretical calculations of the stripe periodicity, we predict that nanometer-scale stripe patterns should be common near two-dimensional critical points.

\begin{acknowledgments}
We thank Thomas Duden and Ernst Bauer for fruitful discussions.
This research was supported by the Office of Basic Energy Sciences, Division
of Materials Sciences, U. S. Department of Energy under Contract No.
DE-AC04-94AL85000, by the Spanish Ministry of Science and Technology
through Project No.~MAT2006-13149-C02-02 and by the Comunidad Aut\'{o}noma 
de Madrid and the Universidad Aut\'{o}noma de Madrid through 
Project No. CCG06-UAM/MAT-0364.
\end{acknowledgments}


\begin{thebibliography}{10}
\bibitem{FFT}
We rotated the
images so the stripes were vertical and removed the background by
dividing by an unfocused image. We then Fourier transformed the center 256
pixels$\times $256 pixels and averaged a minimum of 30 images
in Fourier space. We then extracted the
periodicity and amplitude of the pattern by
fitting the 1D peak along $k_x$ in
Fourier space close to $k_y$=0 to a Lorentzian function.
\bibitem{amplitude}
 A similar analysis for the amplitude is problematic because thermal fluctuations in the stripe position decrease and broaden the Fourier peaks. Thus, the measured amplitude of the peaks is not simply the long-range order of the density oscillations, which would be predicted to vanish as $(T_c-T)^\frac{1}{2}$ rather than the observed $(T_c-T)$.
\bibitem{abinitio}
 W(110) surfaces are
represented by eight layer slabs with up to 18 atoms per layer. For each coverage, to account
for the configurational entropy several possible configurations (up to six
at some coverages) were averaged with each configuration weighted by the
Boltzmann factor, assuming a measurement temperature of 540$^{\circ }$C.
\end{thebibliography}

\end{document}